\documentclass[aps,prl,twocolumn,superscriptaddress,showpacs]{revtex4}

\usepackage{graphicx}
\usepackage{bm}
\usepackage{amssymb}

\begin{document}

\title{\textbf{Semiclassical Dynamics of Electron Wave Packet States with Phase Vortices}}

\author{Konstantin Yu. Bliokh}

\affiliation{Frontier Research System, The Institute of Physical and Chemical Research (RIKEN),
Wako-shi, Saitama 351-0198, Japan}

\affiliation{Institute of Radio Astronomy, 4 Krasnoznamyonnaya st., Kharkov 61002, Ukraine }

\author{Yury P. Bliokh}

\affiliation{Frontier Research System, The Institute of Physical and Chemical Research (RIKEN),
Wako-shi, Saitama 351-0198, Japan}

\affiliation{Physics Department, Technion-Israel Institute of Technology, Haifa 32000, Israel}

\author{Sergey Savel'ev}

\affiliation{Frontier Research System, The Institute of Physical and Chemical Research (RIKEN),
Wako-shi, Saitama 351-0198, Japan}

\affiliation{ Department of Physics, Loughborough University, Loughborough LE11 3TU, United
Kingdom}

\author{Franco Nori}

\affiliation{Frontier Research System, The Institute of Physical and Chemical Research
(RIKEN), Wako-shi, Saitama 351-0198, Japan}

\affiliation{
%Center for Theoretical Physics,
Department of Physics, CSCS, University of Michigan,
Ann Arbor, Michigan 48109-1040, USA}

\begin{abstract}
We consider semiclassical higher-order wave packet solutions of the Schr\"{o}dinger
equation with phase vortices. The vortex line is aligned with the propagation direction,
and the wave packet carries a well-defined orbital angular momentum (OAM) $\hbar l$ ($l$
is the vortex strength) along its main linear momentum. The probability current coils
around momentum in such OAM states of electrons. In an electric field, these states
evolve like massless particles with spin $l$. The magnetic-monopole Berry curvature
appears in momentum space, which results in a spin-orbit-type interaction and a
Berry/Magnus transverse force acting on the wave packet. This brings about the OAM Hall
effect. In a magnetic field, there is a Zeeman interaction, which, can lead to more
complicated dynamics.
\end{abstract}

\pacs{03.65.Vf, 03.65.Sq, 03.75.-b, 72.10.-d} \maketitle

\emph{Introduction.---} The phase front singularities of a wave field, vortices, and
related non-integrable phases have been introduced and examined in seminal papers [1--5].
Phase vortices appear naturally in the electron eigenstates in atoms, quantum Hall
fluids, supermedia, ferromagnets, Bose-Einstein condensates, and classical wave fields
(e.g., in optics). While 2D vortices in condensed matter physics are point-like objects,
with vorticity being orthogonal to the plane of motion [6], the optical vortices are
mainly considered as linear objects in 3D space, with vorticity being aligned with the
wave momentum. Wave beams with vortices constitute a fundamental set of modes with
well-defined orbital angular momentum (OAM) [5]. Recently, these beams have found
numerous applications both in classical and quantum optics [3,5].

Simultaneously, topological phenomena related to the semiclassical wave-packet dynamics
of quantum particles have motivated intensive investigations in various areas of physics:
condensed matter, high energy physics, optics, etc. [6--12]. This has resulted in the
revisiting of the semiclassical equations of motion, but now with the Berry-phase terms
taken into account, and the discovery of such phenomena as the spin Hall effect, with
potential applications. In most cases, the topological Berry terms are due to the
(pseudo)spin of particles, whereas the particle's phase front is implicitly assumed to be
locally of the plane-wave-type, i.e. without singularities.

The dynamics of various types of vortex states with phase singularities also undergo the
action of the topological Berry force, which can be associated with the Magnus force
[13--15]. Both vortex and spin dynamics relate to such fundamental concepts as magnetic
monopoles, Berry phase, space non-commutativity, and generalized Hamiltonian dynamics
[1,2,4,6--16]. It has been shown very recently that the dynamics of optical beams with
vortices reveals all the topological phenomena previously associated with spin, but now
they are related to the intrinsic OAM carried by the vortex [15,17,18]. In contrast to
spin effects, these phenomena are polarization-independent and can be much larger in
magnitude, since the OAM (the vortex strength) can take arbitrarily large values [15].
The OAM Hall effect (an analogue of the spin Hall effect) has also drawn attention in
semiconductor physics [19], but there it is also assumed that the wave packet states of
the electrons have \emph{zero intrinsic} OAM.

While confined electron states with vortices are well-studied in molecules and 2D
structures [20], here we aim to analyze the semiclassical dynamics of a 3D propagating
electron wave packet or beam with intrinsic OAM due to a phase vortex. For simplicity, we
will consider non-relativistic scalar electrons without spin. As we will show, electron
wave packets with OAM manifest fundamental topological and dynamical features which were
associated so far mostly with spin dynamics.

\emph{OAM states of a free electron.---} The Schr\"{o}dinger equation in free space reads
$\left( {i\hbar \frac{\partial }{{\partial t}} + \frac{{\hbar ^2 }}{{2m}}\nabla ^2 }
\right)\psi = 0$. Let us construct semiclassical (paraxial) wave packet solutions
propagating along the $z$ axis and characterized by a narrow distribution in $\left(
{{\bf{p}},E} \right)$ space around some center at $\left( {{\bf{p}}_{\rm c} ,E_{\rm
c}}\right)=\left(p_{\rm c}{\bf e}_z, p_{\rm c}^2 /2m \right)$. Usually, one assumes
Gaussian-type wave packets with the maximal probability density and nearly-plane phase
front in its center [8]. However, this is not the case for higher-order modes. By making
the ansatz $\psi = \exp \left[ {i\hbar ^{ - 1} \left( {p_{\rm c} z - E_{\rm c} t}
\right)} \right] u$, where $u$ is a smooth function with respect to $z$ and $t$, and
neglecting the second-order derivative $\partial ^2 u/\partial z^2$, we arrive at the
parabolic-type equation
\begin{equation}\label{eq1}
\left( {i\hbar \frac{\partial }{{\partial \tau }} + \frac{{\hbar ^2 }}{{2m}}\nabla _ \bot
^2 } \right) u = 0~.
\end{equation}
Here $\nabla _ \bot ^2  = \nabla _x^2  + \nabla _y^2$ and $\partial /\partial \tau  =
\partial /\partial t + \left( {p_{\mathop{\rm c}\nolimits}  /m} \right)\partial /\partial
z$ is the time derivative in the coordinate frame $\left( {x,y,\zeta ,\tau } \right)$
moving with the wave packet center ($\zeta  = z - p_{\rm c}t/m$, $\tau  = t$). Since the
operator in Eq. (1) is $\zeta$-independent, such modes allow factorization, so that one
can choose the transverse and longitudinal parts in the form of the known in optics
Laguerre--Gaussian (LG) beams (with an azimuthally-symmetric intensity profile) and
Hermite--Gaussian (HG) wave packets:
\begin{equation}\label{eq2}
u_{l,{\sf m},{\sf n}} \left(r,\varphi,\zeta ,\tau\right) = u_{l,{\sf m}}^{LG}
\left(r,\varphi,\tau\right) u_{\sf n}^{HG} \left( \zeta \right)~.
\end{equation}
Here $\left( {r,\varphi } \right)$ are polar coordinates in the $\left( {x,y} \right)$
plane, whereas $l = 0, \pm 1, \pm 2,...$ and ${\sf m},{\sf n} = 0,1,2,...$ are the
quantum numbers corresponding to the azymuthal, radial, and longitudinal directions. The
zeroth mode, $l={\sf m}={\sf n}=0$, is a usual Gaussian wave packet.

The explicit form of standard LG and HG solutions can be found in the optics literature
[5], while here we will only emphasize their most significant features. First, they
represent wave packets with Gaussian envelopes and constitute a complete orthonormal set
of modes, so that $\left\langle {u_{l,{\sf m},{\sf n}} } \right.\left| {u_{l',{\sf
m}',{\sf n}'} } \right\rangle = \delta _{ll'} \delta_{{\sf mm}'}\delta_{{\sf nn}'}$, and
any localized wave function can be represented as a superposition $\left| u \right\rangle
= \sum\limits_{l,{\sf m},{\sf n}} {a_{l,{\sf m},{\sf n}} \left| {u_{l,{\sf m},{\sf n}}}
\right\rangle}$. Second, LG modes with $l \ne 0$ contain a screw dislocation of the phase
front on the wave packet axis, $u_{l,{\sf m}}^{LG}\propto\exp(il\varphi)$; in other
words, they contain a \textit{phase vortex} of strength $l$ at $r=0$. Owing to this, the
solutions (2) have a well-defined $z$-component of the OAM. Indeed, $\hat L_z \left|
{u_{l,{\sf m},{\sf n}} } \right\rangle  = \hbar l\left| {u_{l,{\sf m},{\sf n}}}
\right\rangle$ ($\hat L_z =  - i\hbar\partial /\partial \varphi$), and, effectively, the
electron possesses an \textit{intrinsic angular momentum} ${\bf L}\equiv \hbar{\bf
l}=\hbar l {\bf e}_z$. The wave packets (2) also have a \textit{magnetic moment}
$\bm{\mu}=g\mu_B {\bf l}$ ($\mu_B =e\hbar/2m$, $e=-|e|$, $c=1$), where $g=1$ for
classical orbital motion, but the $g$-factor can be different in general (e.g., $g=2$ for
electron spin).

The transverse distribution of the probability density, $\rho=|u|^2$, for LG modes with
$l\ne0$, represents $({\sf m}+1)$ concentric circles and vanishes at $r=0$. At the same
time, the probability current coils around $z$: ${\bf{j}} = m^{-1}[\rho{\bf{p}}_{\rm
c}+\hbar{\rm Im}(u^{*}\nabla u)] \simeq m^{-1}\rho\left({\bf{p}}_{\rm c} + \hbar
l{\bf{e}}_\varphi /r\right)$, Fig.~1. (The same behavior is characteristic for the
Pointing vector of optical LG beams [5].) This implies that the electron trajectories are
effectively \textit{spiral} in \emph{free} space. This effect disappears in the classical
limit $\hbar=0$, and can be regarded as a \textit{zitterbewegung} due to the intrinsic
OAM. In what follows, we will consider only trajectories of the \emph{center} of the wave
packet (2), which is a sort of guiding center, propagating rectilinearly in free space.
\begin{figure}[tbh]
\centering \scalebox{0.4}{\includegraphics{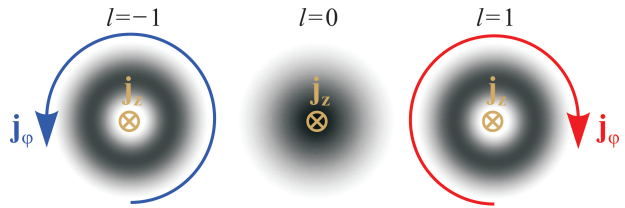}} \caption{(Color online.)
Transverse distribution of the probability density $\rho$ in LG beams with ${\sf m}=0$
and different values of OAM, $l$. Shown are the directions of common $z$ component and
different $\varphi$ components of the probability current ${\bf j}$.} \label{Fig1}
\end{figure}

\emph{Dynamics in external fields.---} In an external electromagnetic potential
$\left({\bf A},\Phi \right)$ generating field $\left( {\bf E},{\bf B}\right)$, the
Schr\"{o}dinger equation reads $\left[ {i\hbar \frac{\partial }{{\partial t}} - e\Phi  -
\frac{1}{{2m}}\left( { - i\hbar \nabla - e{\bf{A}}} \right)^2 } \right] \psi = 0$. The
potentials are assumed to be smoothly space- and time-dependent (the fields are weak) to
ensure the independent adiabatic evolution of the modes (2). We assume that the
wave-packet center, $({\bf p}_{\rm c},{\bf r}_{\rm c})$ (in what follows, the subscripts
``c'' are omited), moves in the vicinity of classical electron trajectory, ${\bf{\dot p}}
= e{\bf{E}} + e{\bf{\dot r}} \times {\bf{B}}$, ${\bf{\dot r}} = {\bf p}/m$, and
approximately conserves its form (2) in the accompanying coordinate system with the $z$
axis locally directed along ${\bf p}$ ($\zeta=z-\int pdt/m$). This means that,
effectively, we deal with a `relativistic' configuration where ${\bf l}=l{\bf p}/p$ and
the `helicity', ${\bf lp}/p=l$, is conserved (see below). The local coordinate frame is
transported along a curved trajectory and. This produces a phase shift due to the Berry
phase [4] and a deflection of the wave packet center due to the geometrical force
[6--12,15], both described by the effective Berry gauge potential (connection) and field
(curvature).

Geometrically, the Berry connection provides for the parallel transport of the state
vector over the phase space [4,7]. Since the $z$ axis is attached now to $\textbf{p}$,
the solution (2) becomes essentially momentum-dependent. The parallel transport of the
wave packet over momentum space implies the covariant derivative $D/D{\bf{p}} =
\partial /\partial {\bf{p}} + {\bm{\mathcal
A}}^{(l)} \left( {{\bf p}} \right)$, where $\bm{\mathcal A}^{(l)} = i\left\langle
{u_{l,{\sf m},{\sf n}}} \right|\partial /\partial {\bf p}\left| {u_{l,{\sf m},{\sf n}}}
\right\rangle$ is the Berry connection, whereas the corresponding curvature is
$\bm{\mathcal B}^{(l)} =
\partial /\partial {\bf p} \times \bm{\mathcal A}^{(l)}$. As will be seen, only the quantum
index $l$ noticeably contributes to the Berry connection. Furthermore, $\left\langle
{u_{l,{\sf m},{\sf n}}} \right|\partial /\partial {\bf p}\left| {u_{l',{\sf m}',{\sf
n}'}} \right\rangle\propto \delta _{ll'} \delta _{{\sf mm}'} \delta _{{\sf nn}'}$, which
evidences the independent evolution of the modes (2). As in the relativistic case of
massless particles with spin, the Berry gauge field takes the form of a `magnetic
monopole', which originates from the local vortex structure $\exp \left( {il\varphi }
\right)$ in the wave packet [15]. Indeed, $\exp \left( {il\varphi } \right) = \left(
{e_{x} + ie_{y} } \right)^l$, where ${\bf{e}}$ is a unit vector orthogonal to ${\bf p}$.
Under variations of ${\bf p}$, ${\bf{e}}$ moves on the unit sphere ${\bf p}/p$, which
leads to the magnetic-monopole-type connection $\bm{\mathcal A} = i\left( {e_{x}  -
ie_{y} } \right)\partial /\partial {\bf p}\left( {e_{x}  + ie_{y} } \right)$ and
corresponding curvature $\bm{\mathcal B} =  - {\bf p}/p^3$ [4,11,12]. As a result we have
$\bm{\mathcal A}^{(l)} = l\bm{\mathcal A}$ and $\bm{\mathcal B}^{(l)} = l\bm{\mathcal
B}$, so that each mode is characterized by the `charge' $l$ in the `magnetic monopole'
field in the momentum space.

The semiclassical dynamics of the wave-packet center can be described involving the
minimal coupling prescription for the electromagnetic and Berry's gauge fields. This
results in the Lagrangian [8]
\begin{equation}\label{eq3}
{\mathcal L} = - {\mathcal H} + {\bf p\dot r} + e{\bf{A\dot r}} + \hbar l{\bm{\mathcal
A}\bf\dot p}~.
\end{equation}
Here the Hamiltonian acquires the energy correction due to the \textit{Zeeman
interaction} of the intrinsic OAM with the magnetic field [8,9,12]: ${\mathcal H} =
\frac{p^2}{{2m}} + e\Phi  + \Delta = E$, $\Delta = -{\bm \mu}{\bf B}$, where the magnetic
moment ${\bm\mu}=g\mu_B {\bf l}$ is also momentum-dependent now. The Berry phase term,
last in Eq. (3), is of the form of the relativistic spin-orbit interaction [12], but now
this is the \textit{orbit-orbit interaction} between the intrinsic OAM and the external
degrees of freedom. Eq.~(3) yields the common phase of the wave packet :
\begin{equation}\label{eq4}
\theta  = \hbar ^{ - 1} \int {\left( {{\bf p}d{\bf r} - Edt} \right)}  + \hbar ^{ - 1}
e\int {\bf{A}} d{\bf r} + l\int \bm{\mathcal A} d{\bf p}~,
\end{equation}
which substitutes for the free-space one, $\hbar ^{ - 1} \left( pz-Et \right)$. The three
terms in Eq.~(4) are, respectively, the dynamical phase, the Dirac (Aharonov--Bohm) phase
[1,2], and the Berry phase [4,7,8]. The latter provides for the parallel transport of the
transverse structure of the wave packet along the curved trajectory [15,17].

Considering the Euler-Lagrange equations for ${\mathcal L}={\mathcal L}({\bf p},{\bf \dot
p},{\bf r},{\bf \dot r})$, Eq.~(3), we arrive at the semiclassical equations of motion
for the wave-packet center [6--16]:
\begin{equation}\label{eq5}
{\bf \dot p} = e{\bf E} - \frac{{\partial \Delta}}{{\partial {\bf r}}} + e{\bf \dot r}
\times {\bf B}~,~~{\bf \dot r} = \frac{{\bf p}}{m} + \frac{{\partial \Delta}}{{\partial
{\bf p}}} - \hbar l{\bf \dot p} \times \bm{\mathcal B}~.
\end{equation}
These equations follow from the Hamiltonian formalism as well, where the minimal coupling
with the electromagnetic and Berry's fields implies a deformation of the symplectic
structure $\Omega = g^{ij} dX_i \wedge dX_j/2$ [7,16]:
\begin{equation}\label{eq6}
\Omega  = {dp_i  \wedge dr_i  + e\frac{\epsilon _{ijk}}{2} B_k dr_i  \wedge dr_j + \hbar
l\frac{\epsilon _{ijk}}{2} {\cal B}_k dp_i \wedge dp_j }~.
\end{equation}
Here ${\bf{X}} = \left( {{\bf{r}},{\bf{p}}} \right)$, and $g^{ij}$ is the symplectic
metric tensor. The equations of motion (5) take the simple form $g^{ij} \dot X_j =
\partial {\mathcal H}/\partial X_i$, whereas the Poisson brackets of the dynamical variables
(or commutators of the corresponding operators) become non-trivial, $\left\{ X_i ,X_j
\right\}= g_{ij}$ (cf. [7,12,16,21]):
\begin{eqnarray}\label{eq7}
\left\{p_i,p_j \right\} & = & D^{ - 1}e\epsilon _{ijk} B_k~,~~
\left\{ r_i,r_j \right\} = D^{ - 1} \hbar l\epsilon _{ijk} {\cal B}_k~,\nonumber\\
\left\{ r_i,p_j \right\} & = & D^{ - 1} \left(\delta _{ij}-e\hbar l B_i
\mathcal{B}_j\right)~,~~D = \sqrt {\det g^{ij}}~.
\end{eqnarray}
Here $D = 1 - e\hbar l{\bf B}\bm{\mathcal B}$ is a correction to the phase space volume
which modifies the density of states [21].

Let us briefly analyze the most remarkable features of equations (5). The last term in
the second equation (5), given by $\hbar l{\bf \dot p} \times {\bf p}/p^3$, represents
the anomalous velocity (${\bf \dot r} \nparallel {\bf p}$) or the `Lorentz force' from
the Berry's `magnetic monopole' in momentum space [6--12]. It causes the transverse
deflection of the wave packet trajectory, which is proportional to the intrinsic OAM
$\hbar l$ [15]. This results in the transverse orbital current, or \textit{OAM Hall
effect}, similar to the spin current in the spin-Hall effect [10--12]. For instance, when
${\bf B}=0$, ${\bf E}={\rm const}$, Eqs.~(5) take the simple form: ${\bf \dot p} = e{\bf
E}$, ${\bf \dot r} = {\bf p}/m + e\hbar l{\bf E} \times {\bf p}/p^3$, and can be readily
integrated analytically (see [10]). They show the transverse shift of the trajectories,
equaling $\hbar l/p_0$ when ${\bf p}_0\bot{\bf E}$, Fig.~2. The OAM Hall effect can be
much stronger than the spin one, since it is proportional to $l$, which formally can take
arbitrarily large values. At the same time, the anomalous velocity is a counterpart of
the Magnus force for 2D vortices which also appears due to the Berry curvature term, but
in coordinate rather than momentum space [13,14].  In an external magnetic field, ${\bf
E}=0$, ${\bf B}={\rm const}$ (for simplicity, let ${\bf B}\perp{\bf p}$), Eqs.~(5) are
reduced, in the linear approximation in $\hbar$, to ${\bf \dot p} = \frac{e}{m}{\bf p}
\times {\bf B}$, ${\bf \dot r} = m^{-1}[{\bf p}+{e\hbar l}(1-\frac{g}{2}){\bf B}/p]$.
Thus, at $g\neq 2$, an OAM-dependent transport of electrons can appear in a constant
magnetic field.
\begin{figure}[tbh]
\centering \scalebox{0.4}{\includegraphics{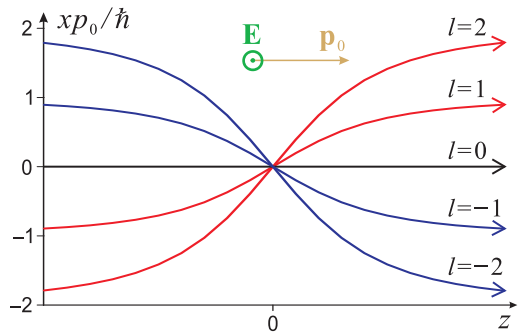}} \caption{(Color online.) Central
trajectories, Eqs.~5, of the electron states with different values of OAM, $l$, moving in
a uniform electric field ${\bf E}=E{\bf e}_y$ (${\bf p}_0=p_0{\bf e}_z$, ${\bf r}_0=0$,
and $l=0$ line corresponds to the classical trajectory).} \label{Fig2}
\end{figure}

The intrinsic OAM, ${\bf l}$, can also be considered as an independent intrinsic
dynamical variable. It can be shown that the helicity ${\bf lp}/p$ is conserved during
the evolution, Eqs.~(5), if ${\bf l}$ obeys the equation
\begin{equation}\label{eq8}
{\bf \dot l} = -\left(\frac{e}{m}{\bf B}+\frac{e{\bf E}\times{\bf
p}}{p^2}\right)\times{\bf l}~.
\end{equation}
Eq.~(8) is of the form of the well-known BMT equation for the electron spin precession
[22,12], where the Berry phase (spin-orbit) term is taken in the relativistic limit
(because intrinsic OAM reveals topological features similar to the spin of a massless
particle), while all other terms are non-relativistic. Eq.~(8) is compatible with
Lagrangian (3) only if either $g=2$ or ${\bf B}=0$. Under evolution of the OAM with
$g\neq 2$ in a magnetic field, ${\bf l}$ does not follow ${\bf p}$ (the precession
frequency of OAM differs from the cyclotron frequency), and thus the initial assumption
of this study is not valid. In this case, the problem becomes much more complicated and
requires an independent study.

\emph{Discussion. Comparison with optics.---} Electron states with vortices can appear in
both bulk solids and in free space. These may open new avenues of research in condensed
matter physics (OAM Hall effect) as well as in electron microscopy and holography [23]
(similarly to singular optics). A comparative analysis of the electron OAM states and
their optical counterparts is summarized in the Table I. Dynamics of optical beams with
vortices is described by equations similar to Eqs.~(5), where the refractive index of the
medium, $n$, plays the role of the external scalar potential $\Phi$ [15]. At the same
time, there is no counterpart of the external curl potential and magnetic field in
optics. Therefore, all electric-field-related features are absolutely similar in electron
and optical systems, while the magnetic-field-related effects are inherent to the
evolution of charged particles only [24].

It is important to find some ways for creating electron phase vortices. On the one hand,
the phase vortex \emph{cannot} be created from a plane wave by a large-scale
electromagnetic potential. Indeed, the phase incursion at the closed contour around the
vortex line equals $2\pi l$. A Dirac phase of this kind can appear only due to a real
magnetic monopole [1,23] or in the presence of an infinite Aharonov--Bohm solenoid [2].
On the other hand, a phase vortex can be created by a short-scale potential. For
instance, an optical vortex can be produced using a diffracting grating with an edge
dislocation (`fork'). For electrons, such a grating can be provided by a real thin
crystal plate with a dislocation. This offers the opportunity of using phase vortices in
electron or neutron crystallography. One can also use a bulk solid as an effective
refractive medium for free electrons, because the electron changes its effective mass
there [23]. Then, a spiral-thickness solid plate will create an electron phase vortex
similarly to the analogous optical lens [3,5].

\begin{widetext}
%\begin{tabular}{|p{3cm}||p{6cm}|p{6cm}|}
\begin{tabular}{|l||c|c|}
\hline
 & Optics & Electrons \\
\hline
\hline
wave field & $\textbf{E}$ & $\psi$ \\
\hline
external potential field & $n$, $\nabla n$ & $\Phi$, $\textbf{E}$\\
\hline
external curl field & --- & $\textbf{A}$, $\textbf{B}$ \\
\hline
generation of & hologram (grating with dislocation) & crystal plate with dislocation\\
phase vortex & spiral-thickness lens & spiral-thickness plate \\
 & --- & magnetic monopole \\
\hline common features & \multicolumn{2}{c}{orbit-orbit interaction: Berry phase and
Magnus/Berry force}\\
\hline distinctive features & \multicolumn{2}{c}{Lorentz force, Dirac phase, Zeeman interaction [24], modified density of states}\\
\hline
\end{tabular}
\begin{center}
Table I: Comparison between electron OAM states and their optical counterparts.
\end{center}
\end{widetext}

\end{document}